\documentstyle[11pt,paspconf,psfig]{article}

\begin{document}

\title{Halo Kinematics}

\author{Terry Bridges}
\affil{Institute of Astronomy, Madingley Road, Cambridge, UK, CB3 0HA}






\begin{abstract}

I summarize recent observations of 
the kinematics of hot tracers in elliptical galaxy halos
(globular clusters, planetary
nebulae, and integrated stellar light), and what these tell us about
the dynamics, dark matter content, and formation of ellipticals. 
A generic result is the ubiquity of dark matter halos in ellipticals.
Studies of globular clusters and planetary nebulae are now finding
outer-halo rotation in many ellipticals, with \linebreak
V/$\sigma$ $\simeq$ 1 beyond a few R$_e$.  In some 
giant ellipticals (M49, M87), there are possible kinematic
differences between metal-poor and metal-rich globular clusters.
These results are consistent with a merger origin for ellipticals.
High-quality data and new modelling techniques now make it
possible to determine {\it simultaneously} the orbital anisotropy and
gravitational potential in ellipticals from integrated-light
measurements; such studies now provide the best evidence for dark
matter halos in ellipticals.  
The new generation of 8--10m telescopes, with
multi-object and integral-field spectrographs, will
dramatically increase sample sizes of discrete tracers and provide
two-dimensional spectroscopy of elliptical halos.  New methods of
analysis
will allow robust determinations 
of stellar kinematics and dark matter distributions in a much
larger number of ellipticals.  Comparison with 
numerical simulations, which are becoming ever more detailed
and physically realistic, will become increasingly important.

\end{abstract}




\section{Introduction and Motivation}

{\it Why Study the Halo Kinematics of Elliptical
Galaxies?}

\medskip

\noindent There are several important subjects that we can address via the
spectroscopy of hot tracers in elliptical galaxy halos:  

\begin{itemize}

\item {\it The  amount and distribution of dark matter
in elliptical halos}.\\  Recent N-body simulations of the formation of
galaxy halos in hierarchical clustering scenarios
(e.g. Navarro et al. 1997) predict that elliptical
galaxies have dark matter halos
with a mildly cusped universal density profile.  It is obviously
important to check this observationally. 

\item {\it The kinematics (ie. velocity distributions) of 
elliptical halo populations}.   We would like to determine the
rotation, velocity dispersion, and velocity anisotropy 
for different halo populations, as far into the halo as possible. 

\item {\it Testing of formation models for elliptical galaxies}. \\
In principle, the 
different formation models make different predictions for the kinematics
of halo populations.
In this review I will concentrate
on two main classes of formation models: monolithic/multiphase collapse,
and mergers (hierarchical or otherwise).

\end{itemize}

\noindent{\it Why is it Difficult?}

\medskip

Spiral galaxies conveniently have excellent kinematical tracers,
in the form of rotating disks of neutral and ionized gas, extending
over a wide range of radii. 
In contrast, we know much less about the 
dark matter content and dynamics of elliptical galaxies. 
Masses have been measured from X-ray observations or HI ring velocities
for some ellipticals, and for others various methods described in this
review have allowed us to rule out constant M/L ratios.  However, the
detailed radial mass {\it distributions} of elliptical galaxies are 
almost totally unknown.  

The problem is a fundamental degeneracy between {\it stellar orbits}
and the underlying {\it gravitational potential}.  Stellar test particles
can occupy a wide range of different orbits in a given potential, and
the degeneracy cannot be removed from rotation and velocity dispersion
data alone (e.g. a rising velocity dispersion could be due to either
tangential anisotropy or a dark matter halo).  Only 
recently have high-quality data and new modelling techniques 
allowed the simultaneous determination of the anisotropy
and mass distribution from measuring the {\it shape} of the
line-of-sight velocity distribution (LOSVD) (Section 4).  

\medskip


\noindent{\it Outline of This Review}

\medskip

In this paper I will concentrate on recent observational studies
of the kinematics and dark matter content of elliptical galaxy halos,
with some discussion of the implications for the formation of
elliptical galaxies.
This is not intended to be a complete review of the field, and I 
will point the reader to earlier 
reviews.  I will not discuss the {\it shapes} of dark matter halos
(see review by P. Sackett in this volume), nor the abundance
information that
can also be obtained from halo spectroscopy (see for instance the 
review by R. Bender in this volume).  
In Section 2, I will
discuss planetary nebulae; in Section 3, globular clusters; and in
Section 4, integrated-light measurements.  
In Section 5 I present my main conclusions and predictions/hopes 
for the future.

\section{Planetary Nebulae}

Here I will concentrate on results obtained in the last five years.
For an excellent review of earlier work, 
see Arnaboldi \& Freeman (1997).

\medskip

\subsection{The Usefulness of Planetary Nebulae}

Planetary nebulae (PNe) 
are excellent dynamical probes because:

\begin{itemize}

\item They are bright; $\sim$ 15\% of their energy goes into a single
emission line of [OIII] at 5007 \AA.

\item They are easy to detect: one `simply' compares
an image taken with a narrow-band filter centered on [OIII] with one
taken off-band.  Once PNe have been identified, multi-object spectroscopy
with high dispersion allows velocities to be obtained with a precision
better than 50 km/sec.  Very clean samples are obtained in this way,
with very little contamination. 

\item They are numerous, even in early-type galaxies, and hundreds
of PNe can be measured with 4m class telescopes in galaxies out to
the distance of Virgo and Fornax.  With 8--10m class telescopes,
as many as 500 PNe can be measured in the brightest galaxies in
these clusters.

\end{itemize}

\subsection{Observational Summary}

Table 1 gives an overview of recent dynamical studies  
of ellipticals using PNe.  Column 3 gives the number
of PNe with measured velocities, Column 4 the galactocentric
radius of the most distant measured PNe, Column 5 the
M/L ratio in the B band in solar units, and Column 6
the PNe rotation amplitude.

\begin{table}
\caption{Recent Dynamical Studies of Ellipticals using PNe}
\begin{center}\scriptsize
\begin{tabular}{clcccc}
Galaxy & ~~~~~~Authors &  N$_{PNe}$
 & Radius
 & M/L$_B$ & Rotation \\
       &             &            &  (kpc)   &         &  (kms$^{-1}$) \\
\tableline
         &                       &    &    &   &    \\
NGC 3379 & Ciardullo et al. 1993 & 29 & 10 & 7 & -- \\
NGC 1399 & Arnaboldi et al. 1994 & 37 & 16 & $\sim$ 80 & 290 \\
NGC 3384 & Tremblay et al. 1995 & 68 & 7 & $>$ 9 & 125 \\
NGC 4406 & Arnaboldi et al. 1996 & 16 & 19 & 13 & 250 \\
NGC 1316 & Arnaboldi et al. 1998 & 43 & 16 & 8 & 100 \\
NGC 5128 & Peng et al. 1998 & 657 & 40 & 15 & 100 
\end{tabular}
\end{center}
\end{table}

\noindent Some general comments about Table 1 may be made:

\begin{itemize}

\item The power of
PNe for studying elliptical halos is demonstrated by the fact
that the measurements
extend to at least 2 R$_e$ in every galaxy.

\item The M/L ratio is invariably increasing outwards, from
central values $\sim$ 5 to values of typically 10--20 
at larger radius: dark matter halos seem to be a common feature
of elliptical galaxies.  

\item There is almost always detectable
rotation in the PNe: the halos of elliptical galaxies contain a
large amount of angular momentum (see below). 

\end{itemize}

\noindent{\bf Further Discussion}

\medskip

\noindent{\it NGC 1399 and Fornax Intracluster PNe}

\medskip

NGC 1399 is a giant elliptical galaxy situated at the center 
of the nearby Fornax cluster.
Arnaboldi et al. (1994) measured a large rotation
($\sim$ 300 km/sec) for the PNe in NGC 1399,
showing that the total
specific angular momentum J/M of ellipticals can in fact be 
comparable with that of giant spirals, in agreement with 
cosmological simulations of elliptical formation.  Arnaboldi
et al. also suggested that the large M/L ratio of NGC 1399
was due to PNe responding more to the potential of the
{\it Fornax cluster} than to the {\it galaxy itself};
Figure 1 of Kissler-Patig (1998; this volume) 
shows that there is indeed a transition from the
galaxy to cluster, as traced by stars, PNe, globular clusters,
cluster galaxies, and X-ray gas, at progressively larger radius. 
Such intracluster PNe complicate efforts to study cluster
galaxies, but are obviously extremely important
in terms of studying the clusters themselves.
This is an active field, with several ongoing searches in the  
Fornax 
and Virgo clusters.


\medskip

\noindent{\it NGC 1316}

\medskip

\begin{figure}[h]
\vspace{0in}
\psfig{{figure=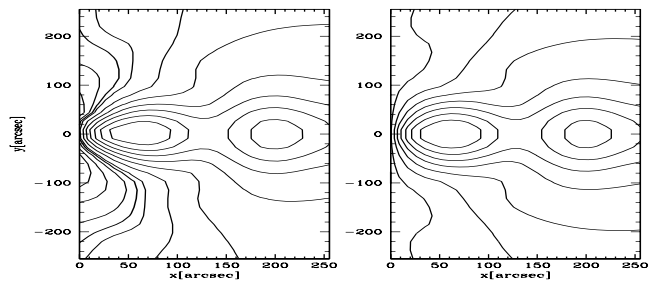},angle=0,height=1.75in,width=3.5in}
\caption{Smoothed velocity field of NGC 1316 obtained from
combining absorption-line and PNe data, taken from Arnaboldi
et al. (1998).  The left panel shows the velocity field including
the minor axis absorption-line velocities; the right panel without.}
\label{fig-1316}
\end{figure}

The work of Arnaboldi et al. (1998, also this volume) 
on NGC 1316 is noteworthy
because it is one of the first attempts to combine PNe and
integrated-light data.  The smoothed velocity field
of the combined PNe and absorption-line data 
(Figure~\ref{fig-1316})  
allows the rotation curve 
in NGC 1316 to be traced from the galaxy center out to $\sim$
16 kpc, and reveals significant outer-halo rotation in this galaxy.
With larger samples of discrete tracers and with the further
addition of X-ray and gravitational lensing data at 
larger radius, such techniques will be
a powerful way to study the dynamics and mass distributions
of ellipticals. 

\medskip

\noindent{\it NGC 5128}

\medskip

\begin{figure}[h]
\vspace{0cm}
\psfig{{figure=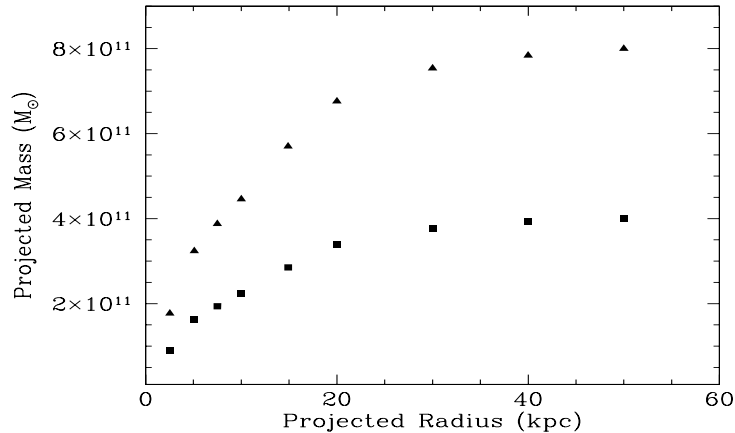},angle=-90,height=6cm,width=10cm}
\caption{Projected mass as a function of radius for NGC 5128
from PNe data of Peng et al. (1998, this volume).  
Squares give the mass
for isotropic orbits, and triangles for radial orbits.}
\label{fig-5128}
\end{figure}

Combined studies by Hui et al. (1995) and 
Peng et al. (1998, this volume) 
of the nearby
peculiar elliptical NGC 5128 (Centaurus A), have amassed the
largest number of PNe velocities in any elliptical and the
best radial coverage.
Hui et al. were able to show that the velocity field 
of the PNe is likely triaxial, and both studies show significant
rotation out to large radius.  
Figure~\ref{fig-5128} shows the
enclosed mass as a function of radius using the Projected Mass
Estimator (PME; Heisler, Tremaine \& Bahcall 1985), taken
from Peng et al.

\section{Globular Clusters}

\subsection{Introduction}

Globular clusters share many of the advantages of PNe: they are
bright (M$_V \simeq -7$) and especially numerous in early-type
galaxies.  However, since they are absorption-line objects, 
long integrations are required (the absorption
lines, however, can also be used to obtain precious {\it abundance}
information).  I will concentrate
on work done since the review by Brodie (1993).

\subsection{Observational Summary}

Table 2 gives an overview of recent dynamical studies of
(mostly!) elliptical galaxies using globular clusters; it
has the same format as Table 1.

\begin{table}
\caption{Recent Studies of Ellipticals using Globular Clusters}
\begin{center}\scriptsize
\begin{tabular}{clcccc}
Galaxy & ~~~~Authors &  N$_{cl}$ & Radius & M/L$_B$ & Rot \\
       &         &            &  (kpc)   &         &  (kms$^{-1}$) \\
\tableline
       &         &            &          &         &      \\
M31 & Perrett et al. 1998 & 220 & 20 & -- & $\sim$ 125 \\
M81 & Perelmuter et al. 1995 & 25 & 19 & 19 & -- \\
M104 & Bridges et al. 1997 & 34 & 14 & 22 & 75 \\
M49 & Zepf et al. 1998 & 144 & 50 & $\sim$ 70 & 100 \\
NGC 3115 & Kavelaars et al. 1998 & 22 & 14 & 20 & 190 \\
NGC 1399 & Grillmair et al. 1994 & 47 & 50 & 70--80 & -- \\
NGC 1399 & Kissler-Patig et al. 1998a & 18 & 28 & 35--75 & -- \\
NGC 1399 & Minniti et al. 1998 & 18 & 28 & 50--130 & -- \\
NGC 1399 & Kissler-Patig et al. 1998b & 74 & 50 & 150--200 & 150??\\
M87 & Cohen \& Ryzhov 1997 & 205 & 40 & 30 & ~100 \\
M87 & Kissler-Patig \& Gebhardt 1998 & 205 & 40 & -- & 300 \\
\end{tabular}
\end{center}
\end{table}

The general comments made about PNe in Section 2.2 also apply to
globular clusters: the data extend out to 5--10 R$_e$, and 
support the existence of dark matter halos in virtually every
elliptical studied.  There is also evidence for outer-halo
rotation in some ellipticals, similar to that found in PNe. 

\medskip

\noindent{\bf Further Discussion}

\medskip

\noindent{\it M87/Virgo and NGC 1399/Fornax}

\medskip

These two centrally-located gE/cD galaxies have been the object
of several studies, as shown in Table 2.
I will only touch on a couple of interesting recent results,
since M87 and NGC 1399 are both reviewed
by Kissler-Patig (1998; this volume).   In NGC 1399, 
Kissler-Patig et al. (1998b) have combined the available
globular cluster velocities, and found evidence
for rotation in the {\it outer-most}
clusters (beyond 5 arcmin radius), an interesting result in
light of the large rotation found in the NGC 1399 PNe by
Arnaboldi et al. (1994) (but in the opposite sense to
the globular clusters!  See Kissler-Patig for more details).  

\begin{figure}[h]
\vspace{0cm}
\psfig{{figure=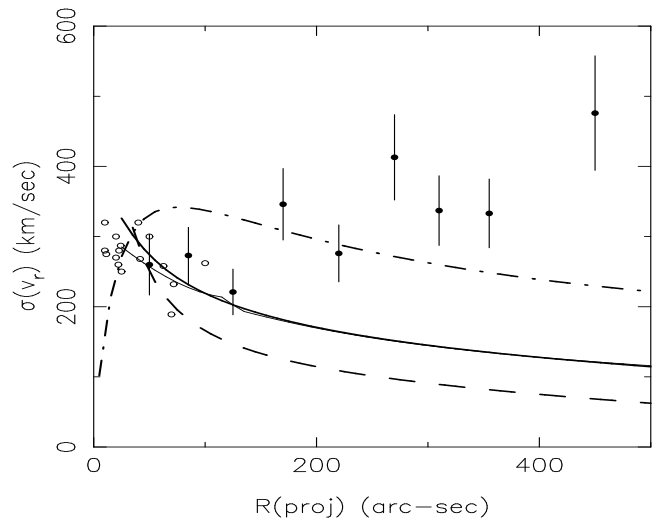},angle=0,height=7cm,width=9cm}
\caption{Velocity dispersion vs. projected radius
for M87, taken from Cohen \& Rhyzov
(1997).  The open circles are for the stars,  
and the filled circles are for the
globular clusters.  The curves are predicted profiles for
various models; see Cohen \& Rhyzov for further details.}
\label{fig-cohen}
\end{figure}

Cohen \& Rhyzov (1997) obtained velocities and spectroscopic
metallicities for 205 globular clusters in M87. 
From Figure~\ref{fig-cohen}, taken from Cohen \& Rhyzov,
we see that the globular cluster
velocity dispersion {\it increases} outwards.  This is 
similar to what is seen in NGC 1399/Fornax, 
and in both cases it is 
likely that we are seeing the transition from the galaxy
to the cluster. 
Kissler-Patig
\& Gebhardt (1998) have re-analyzed the 
Cohen \& Rhyzov data, and found that the {\it metal-poor} clusters
have a higher rotation ($\sim$ 300 km/sec) than
the {\it metal-rich} clusters ($\sim$ 100 km/sec); this is
only a 1-$\sigma$ result, however.

\medskip

\noindent{\it M49/NGC 4472}

\medskip

M49 (=NGC 4472) represents a different environment than M87,
being the central gE in a Virgo southern subcluster.  Thus, we
expect less contamination from any intracluster globular clusters,
and a better determination of the dynamics and dark matter of
M49 itself.   The Washington photometry of 
Geisler et al. (1996) shows that the globular cluster colour
distribution is clearly bimodal, with corresponding metallicity
peaks at 
[Fe/H] = $-$1.3 and $-$0.1.

\begin{figure}[h]
\vspace{0cm}
\psfig{{figure=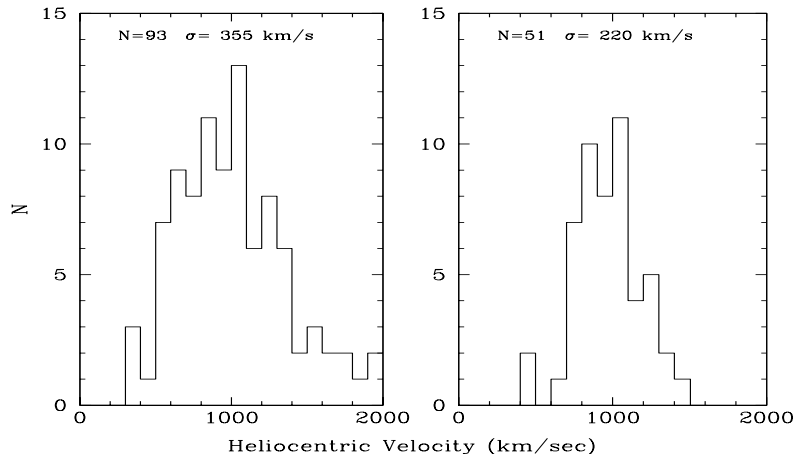},angle=-90,height=6cm,width=11cm}
\caption{Velocity histograms for M49 globular clusters, taken
from Zepf et al. (1998).  {\bf Left:} 
Metal-poor clusters; {\bf Right:} Metal-rich clusters. 
The difference in
velocity dispersion between the metal-poor and metal-rich clusters
is significant at better than 99\%.}
\label{fig-m49sigma}
\end{figure}

We (Sharples et al. 1998) obtained  
velocities for $\sim$ 50 globular clusters in M49 
at the WHT.
We found tentative evidence for kinematic differences between the
red (metal-rich) and blue (metal-poor) globular clusters, 
with the blue clusters having both a higher velocity
dispersion and rotation.  We now have 
144 cluster velocities, with the addition of recent CFHT data 
(Zepf et al. 1998).
Figure~\ref{fig-m49sigma} shows that the  velocity
dispersion difference between the red and blue clusters still holds with
the larger sample, as does the difference in rotation (though the latter
is still only marginally significant).  The larger rotation in the
more extended blue cluster system is consistent with a merger origin
for M49, since mergers are efficient at transporting angular momentum
outwards (e.g. Hernquist \& Bolte 1993).  
As noted above, a similar result has 
been found more recently by Kissler-Patig \& Gebhardt (1998)
for the M87 globular clusters.

\begin{figure}[h]
\psfig{{figure=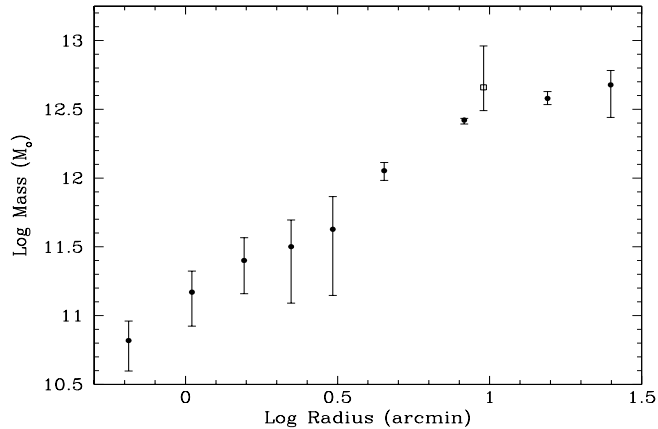},angle=-90,height=6cm,width=9cm}
\caption{Mass determinations for M49.  The open square is the mass determined
from the globular cluster velocities and the Projected Mass Estimator
(Zepf et al. 1998), while
the filled circles are masses derived from ROSAT X-ray data by Irwin \&
Sarazin (1996).}
\label{fig-m49mass}
\end{figure}

We have used the globular cluster velocities and Projected Mass Estimator
(assuming isotropic orbits and an extended mass distribution),
to find a mass of 4.5 $\times$ 10$^{12}$ M$_\odot$ and 
M/L$_B$ ratio of $\sim$ 70 at 50 kpc (6 R$_e$) for M49; these values
are comparable to those found
in M87 and NGC 1399 from similar studies. 
Figure~\ref{fig-m49mass} shows
that the mass as determined from the globular clusters agrees quite well
with that found from ROSAT X-ray data by Irwin \& Sarazin (1996).

\section{Integrated-Light Studies}

\subsection{Introduction}

Studies of the stellar component (integrated-light) 
in elliptical galaxies complement
dynamical studies at larger radius with globular clusters, PNe,
and X-ray  data.  With  integrated-light data, there are not the
same sample size problems as with discrete tracers, but the 
radidly decreasing surface brightness of elliptical halos restricts
such measurements to within $\sim$ 3 R$_e$ of the galaxy center
(at least with 4m-class telescopes).

This field is undergoing a renaissance with recent advances in
observations and analysis techniques.  Observationally, we now
have better CCDs, sky-subtraction techniques, and the careful use
of stars of different spectral types for template matching.  There
is an awareness of the need for long-slit spectroscopy with
several position angles, and now integral-field spectrographs
can provide full two-dimensional coverage, although with
small fields. 

Theoretically, a major step forward has been the development of 
analysis techniques utilizing the {\it entire} Line-of-Sight Velocity
Distribution (LOSVD) (e.g. Rix \& White 1992; van der Marel \& Franx
1993), not just the lowest moments
V and $\sigma$.  With the incorporation of 
these higher-order moments, 
the degeneracy between stellar orbits and gravitational potential can
be removed and in principle both can be determined simultaneously.
There are many possible schemes; see papers by N. Cretton
and S. de Rijcke in this volume for further details.  As I will show
below, recent studies using these techniques on high-quality data
give very strong evidence for dark matter halos in ellipticals.

\subsection{A Selected Review of Recent Work}

This review is far from exhaustive, and is meant only to indicate the
flavour and potential of recent work. 

\medskip

\noindent{\it Saglia et al. 1993}

\medskip

This work represents the beginning of the `modern era' in studies
of integrated light in ellipticals.  Saglia et al. obtained major-axis
spectra out to 2 R$_e$ for NGC 4472, NGC 7144, and IC 4296, 
and found suggestive
evidence that the M/L ratio increases outwards in all three.

\medskip

\noindent{\it Carollo et al. 1995}

\medskip

Carollo et al. obtained major-axis spectra out to 2--2.5 R$_e$ for
NGCs 2434, 2663, 3706, and 5018.  They carried out Gauss-Hermite
modelling of the LOSVDs, and found that strong tangential anisotropy
can be ruled out at large radii: ie., there are very likely 
dark matter halos in all of these galaxies.

\medskip

\noindent{\it Rix et al. 1997}

\medskip

Rix et al. measured the LOSVD out to 2.5 R$_e$ for the E0 galaxy
NGC 2434.  They used a variation of Schwarzschild's 
orbit-building method
(see paper by N. Cretton in this volume)
to rule out constant M/L models, regardless of the orbital
anisotropy. The anisotropy itself is not well-constrained, but the
power of the method for studying dark matter halos is apparent. 

\medskip

\noindent{\it Gerhard et al. 1998}

\medskip

\begin{figure}[h]
\psfig{{figure=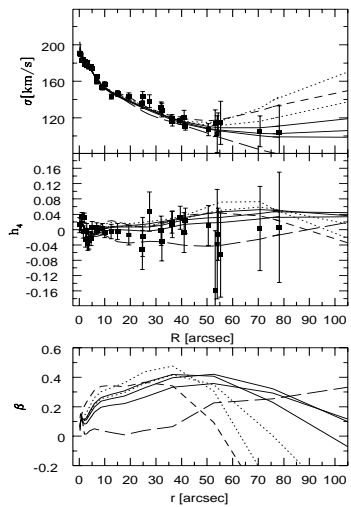},angle=0,height=7cm,width=5cm}
\caption{Dynamical models for NGC 6703 for
various luminous plus dark matter potentials, taken from
Gerhard et al. (1998).  The {\it top} and {\it middle} panels
show fits to $\sigma$ and Gauss-Hermite h$_4$ respectively, 
while the
{\it bottom} panel shows the inferred anisotropy 
$\beta$(r).
See Gerhard et al. 1998 for more details about the various
models.}
\label{fig-6703}
\end{figure}

Gerhard et al. measured the LOSVDs in the E0 galaxy NGC 6703 along
the major axis and parallel to the minor axis.  They used a 
new non-parametric technique to determine the Distribution Function
directly from the kinematic data, avoiding problems associated
with a Gauss-Hermite parameterization.  They showed that both
the M/L ratio and the anisotropy increase outwards: as for 
NGC 2434, no model without dark matter will fit the data. 
Figure~\ref{fig-6703} shows fits to their $\sigma$ and
Gauss-Hermite h$_4$ radial profiles for various
luminous plus dark matter potentials, and the inferred
anisotropy profile $\beta$(r).  

\medskip

\noindent{\it Statler et al. 1996,1998}

\medskip

The strength of the work by Statler et al. is the good
radial coverage (2--3 R$_e$) coupled with spectra taken at four 
position angles.  This allows them to reconstruct the stellar velocity
field, revealing that the galaxy is nearly axisymmetric.  They 
cannot find a constant M/L model (2 or 3 integral) that will fit the
kinematic data, thus providing firm evidence for a dark matter halo.

\medskip

\noindent{\it Carter, Bridges, \& Hau 1998}

\medskip

\begin{figure}[h]
\psfig{{figure=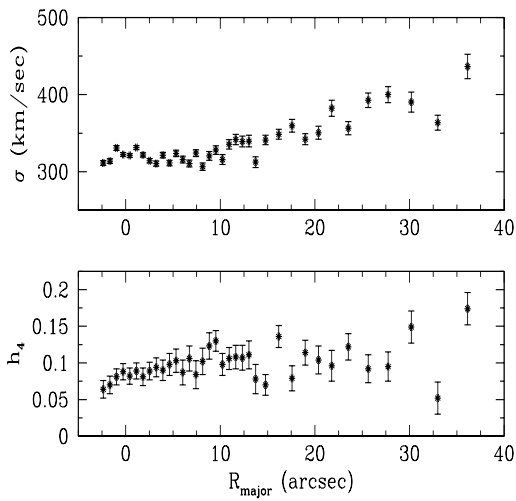},angle=-90,height=7cm,width=7cm}
\caption{Stellar Kinematics in NGC 6166, from
Carter, Bridges, \& Hau (1998).  {\it Upper Panel:} 
Stellar velocity dispersion vs. radius.  {\it Lower Panel:}
Gauss-Hermite parameter h$_4$ vs. radius.} 
\label{fig-6166}
\end{figure}

We have obtained deep major-axis
spectra out to $\sim$ 1 R$_e$ (20 kpc) for 
three cD galaxies (NGCs 6166, 6173, and 6086).  
Gauss-Hermite modelling
of the LOSVDs shows that h$_4$ is constant and
positive in all three galaxies, suggesting constant
radial anisotropy 
(though detailed modelling has
yet to be done).  Thus, the rising velocity dispersion profile 
in NGC 6166 (Figure~\ref{fig-6166}) is almost certainly due to a
dark matter halo, and not tangential anisotropy at large radius. 

\section{Conclusions and the Future}

\subsection{Summary}

\begin{itemize}

\item There is  firm evidence for dark matter 
halos in elliptical 
galaxies, with the M/L ratio increasing from 5--10 at small
radius to $>$ 20 beyond 20 kpc; in some galaxies, M/L ratios
of 50--100 have been found at larger radius. 

\item The evidence for dark matter halos is most secure from 
integrated-light studies, where higher order moments of the
LOSVD can be used to constrain both the orbital anisotropy and
dark matter; in NGC 2434 and NGC 6703 constant M/L models have
been definitively ruled out.   Such studies most often find
stellar radial anisotropy, consistent with recent 
numerical simulations of galaxy formation in
CDM universes (e.g. Dubinksi 1998).

\item Studies using 
PNe and globular clusters are still plagued by
small samples, and assumptions about the anisotropy have to
be made; sample sizes will have to increase by at least an
order of magnitude before truly robust conclusions can be drawn.

\item There is good evidence for outer-halo rotation in many
ellipticals, with V/$\sigma$ $\simeq$ 1 beyond a few R$_e$.
This is consistent with a merger origin for elliptical
galaxies, since hierarchical merging scenarios and
N-body simulations of galaxy mergers both predict 
that ellipticals
have significant angular momentum at large radius. 
Further support for a merger model comes from the finding
that the (more spatially-extended) metal-poor 
globular clusters in M49 and M87 have
a larger rotation than the metal-rich clusters.

\end{itemize}

\subsection{The Future}

On the observational
side, multi-slit
spectrographs are allowing us to obtain useful numbers of 
PNe and globular cluster velocities.
There are 
many ongoing imaging and spectroscopic surveys of
globular clusters and PNe, and increasing numbers
of integrated-light studies, in early-type galaxies
(see several of the papers in this volume).
Integral-field
spectroscopy of elliptical halos
will become increasingly important, especially over
large fields. 
The new generation of 8--10m telescopes with multi-object
spectrographs will dramatically increase sample sizes, and
allow us to study more distant ellipticals.  It will be
especially important to carry out more work on field
ellipticals, which have been rather neglected to date.
With high S/N spectra, we can also hope to obtain
globular cluster
{\it abundances} and {\it ages},
which will tell us more
about the formation epoch and chemical enrichment of ellipticals.
Large telescopes will also allow us
to study the integrated-light out to larger radius, and
we will be better able to compare the stellar kinematics
with globular clusters and PNe.

On the theoretical side, there has been considerable  
recent work on developing methods to extract the 
most information from available 
high-quality integrated-light
data,
with an emphasis on robust, non-parametric methods.  
Larger samples of discrete velocities will require new
methods of analysis, for instance the non-parametric
techniques of Merritt and collaborators
(e.g. Merritt \& Saha 1993); 
with 500--1000 (!) velocities per galaxy,
both the anisotropy and dark matter can be 
determined simultaneously.  In order to obtain a coherent
dynamical picture of elliptical galaxies, we need to 
develop methods that combine {\it all} available kinematic
data, and there have been some recent steps in this
direction (e.g. Arnaboldi et al. 1998).

\acknowledgements

It is a pleasure to thank Magda Arnaboldi, Nicolas Cretton,
Herwig Dejonghe,
John Feldmeier, Ken Freeman, J.J. Kavelaars, Ortwin Gerhard,
Markus Kissler-Patig, Eric Peng, Hans-Walter Rix, and Tom Statler,
for discussions, preprints, and figures.  I'd also like to thank
my collaborators, Keith Ashman, Mike Beasley, Dave Carter, Doug
Geisler, Dave Hanes, Kathy Perrett, and Steve Zepf, for allowing
me to discuss our joint work, and for their support and
encouragement.  Many thanks to the organizers for inviting 
me to this wonderful conference,
and congratulations to the local committee for an excellent job; Monica
Valluri, Dave Merritt, and Stacy McGaugh deserve
special thanks.  I'd like to offer humble
apologies to Alex Turnbull and Jim Collett for
facing the wrath of the police on my account.


%

\end{document}